\documentclass[11pt,a4paper]{article}
\usepackage[hyperref]{emnlp2018}
\usepackage{times}
\usepackage{latexsym}
\usepackage{graphicx}
\usepackage{url}

\aclfinalcopy 

\usepackage{booktabs}
\usepackage{array}
\usepackage{adjustbox}
\usepackage[T1]{fontenc}
\usepackage{float}
\usepackage{algpseudocode}
\usepackage{algorithm}
\usepackage{caption}
\usepackage{amsmath,amssymb}

\usepackage[colorinlistoftodos]{todonotes}

\usepackage{multirow}

\title{Predicting Factuality of Reporting and Bias of News Media Sources}

\author{
	Ramy Baly$^1$, Georgi Karadzhov$^3$, Dimitar Alexandrov$^3$, James Glass$^1$, Preslav Nakov$^2$\\
    $^1$MIT Computer Science and Artificial Intelligence Laboratory, MA, USA\\
    $^2$Qatar Computing Research Institute, HBKU, Qatar;\\
    $^3$Sofia University, Bulgaria\\
    {\tt \{baly, glass\}@mit.edu, pnakov@qf.org.qa}\\
    {\tt \{georgi.m.karadjov, Dimityr.Alexandrov\}@gmail.com}
 }

\date{}

\begin{document}

\maketitle

\begin{abstract}
We present a study on predicting the factuality of reporting and bias of news media.
While previous work has focused on studying the veracity of claims or documents, here we are interested in characterizing entire news media.
These are under-studied but arguably important research problems, both in their own right and as a prior for fact-checking systems.
We experiment with a large list of news websites and with a rich set of features derived from (\emph{i})~a sample of articles from the target news medium,
(\emph{ii})~its Wikipedia page,	
(\emph{iii})~its Twitter account,
(\emph{iv})~the structure of its URL, and
(\emph{v})~information about the Web traffic it attracts.
The experimental results show sizable performance gains over the baselines, and confirm the importance of each feature type.
\end{abstract}

\section{Introduction}
\label{sec:intro}

The rise of social media has democratized content creation and has made it easy for everybody to share and spread information online.
On the positive side, this has given rise to citizen journalism, thus enabling much faster dissemination of information compared to what was possible with newspapers, radio, and TV.
On the negative side, stripping traditional media from their gate-keeping role has left the public unprotected against the spread of misinformation, which could now travel at breaking-news speed over the same democratic channel.
This has given rise to the proliferation of false information that is typically created either
(\emph{a})~to attract network traffic and gain financially from showing online advertisements, e.g.,~as is the case of \emph{clickbait}, or
(\emph{b})~to affect individual people's beliefs, and ultimately to influence major events such as political elections~\cite{Vosoughi1146}.
There are strong indications that false information was weaponized at an unprecedented scale during the 2016 U.S. presidential campaign.

\noindent ``Fake news'', which can be defined as ``fabricated information that mimics news media content in form but not in organizational process or intent''~\cite{Lazer1094},  became the word of the year in 2017, according to Collins Dictionary.
``Fake news'' thrive on social media thanks to the mechanism of sharing, which amplifies effect.
Moreover, it has been shown that ``fake news'' spread faster than real news~\citep{Vosoughi1146}. As they reach the same user several times, the effect is that they are perceived as more credible, unlike old-fashioned spam that typically dies the moment it reaches its recipients.
Naturally, limiting the sharing of ``fake news'' is a major focus for social media such as Facebook and Twitter.

Additional efforts to combat ``fake news'' have been led by fact-checking organizations such as Snopes, FactCheck and Politifact, which manually verify claims.
Unfortunately, this is inefficient for several reasons.
First, manual fact-checking is slow and debunking false information comes too late to have any significant impact. At the same time, automatic fact-checking lags behind in terms of accuracy, and it is generally not trusted by human users. In fact, even when done by reputable fact-checking organizations, debunking does little to convince those who already believe in false information.

A third, and arguably more promising, way to fight ``fake news'' is to focus on their source.
While ``fake news'' are spreading primarily on social media, they still need a ``home'', i.e., a website where they would be posted.
Thus, if a website is known to have published non-factual information in the past, it is likely to do so in the future.
Verifying the reliability of the source of information is one of the basic tools that journalists in traditional media use to verify information.
It is also arguably an important prior for fact-checking systems
\cite{Popat:2017:TLE:3041021.3055133,DBLP:conf/aaai/NguyenKLW18}.

\noindent Fact-checking organizations have been producing lists of unreliable online news sources, but these are incomplete and get outdated quickly.
Therefore, there is a need to predict the factuality of reporting for a given online medium automatically, which is the focus of the present work. We further study the bias of the source (left vs. right), as the two problems are inter-connected, e.g., extreme-left and extreme-right websites tend to score low in terms of factual reporting.
Our contributions can be summarized as follows:

\begin{itemize}
  \setlength{\parskip}{0pt}
  \setlength{\itemsep}{5pt}

\item We focus on an under-explored but arguably very important problem: predicting the factuality of reporting of a news medium. We further study bias, which is also under-explored.

\item We create a new dataset of news media sources, which has annotations for both tasks, and is 1-2 orders of magnitude larger than what was used in previous work. We release the dataset and our code, which should facilitate future research.\footnote{The data and the code are at \url{https://github.com/ramybaly/News-Media-Reliability/}}

\item We use a variety of sources such as
(\emph{i})~a sample of articles from the target website,
(\emph{ii})~its Wikipedia page,
(\emph{iii})~its Twitter account,
(\emph{iv})~the structure of its URL, and
(\emph{v})~information about the Web traffic it has attracted.
This combination, as well as some of the sources, are novel for these problems.

\item We further perform an ablation study of the impact of the individual (groups of) features.

\end{itemize}

The remainder of this paper is organized as follows: Section~\ref{sec:related} provides an overview of related work. Section~\ref{sec:method} describes our method and features. Section~\ref{sec:experiments} presents the data, the experiments, and the evaluation results. Finally, Section~\ref{sec:conclusion} concludes with some directions for future work.

\section{Related Work}\label{sec:related}

Journalists, online users, and researchers are well-aware of the proliferation of false information, and thus topics such as credibility and fact-checking are becoming increasingly important.
For example, the ACM Transactions on Information Systems journal dedicated, in 2016, a special issue on Trust and Veracity of Information in Social Media \cite{Papadopoulos:2016:OSI}.

\noindent There have also been some related shared tasks such as the SemEval-2017 task~8 on Rumor Detection \cite{derczynski-EtAl:2017:SemEval}, the CLEF-2018 lab on Automatic Identification and Verification of Claims in Political Debates \cite{clef2018checkthat:task1, clef2018checkthat:task2,clef2018checkthat:overall}, and the FEVER-2018 task on Fact Extraction and VERification \cite{thorne-EtAl:2018:N18-1}.

The interested reader can learn more about ``fake news'' from the overview by \citet{Shu:2017:FND:3137597.3137600}, which adopted a data mining perspective and focused on social media.
Another recent survey was run by \citet{thorne-vlachos:2018:C18-1}, which took a fact-checking perspective on ``fake news'' and related problems.
Yet another survey was performed by~\citet{Li:2016:STD:2897350.2897352}, covering truth discovery in general.
Moreover, there were two recent articles in \emph{Science}:
\citet{Lazer1094} offered a general overview and discussion on the science of ``fake news'', while
\citet{Vosoughi1146} focused on the process of proliferation of true and false news online.
In particular, they analyzed 126K stories tweeted by 3M people more than 4.5M times, and confirmed that ``fake news'' spread much wider than true news.

Veracity of information has been studied at different levels:
(\emph{i})~claim-level (e.g.,~\emph{fact-checking}),
(\emph{ii})~article-level (e.g.,~\emph{``fake news'' detection}),
(\emph{iii})~user-level (e.g.,~\emph{hunting for trolls}), and
(\emph{iv})~medium-level (e.g.,~\emph{source reliability estimation}).
Our primary interest here is in the latter.

\subsection{Fact-Checking}

At the claim-level, fact-checking and rumor detection have been primarily addressed using information extracted from social media, i.e.,~based on how users comment on the target claim \cite{Canini:2011,Castillo:2011:ICT:1963405.1963500,Ma:2015:DRU,ma2016detecting,PlosONE:2016,P17-1066,dungs-EtAl:2018:C18-1,kochkina-liakata-zubiaga:2018:C18-1}.
The Web has also been used as a source of information \cite{mukherjee2015leveraging,popat2016credibility,Popat:2017:TLE:3041021.3055133,RANLP2017:factchecking:external,AAAI2018:factchecking,baly-EtAl:2018:N18-2}.

In both cases, the most important information sources are
\emph{stance} (does a tweet or a news article agree or disagree with the claim?), and
\emph{source reliability} (do we trust the user who posted the tweet or the medium that published the news article?).
Other important sources are linguistic expression, meta information, and temporal dynamics.

\subsection{Stance Detection}

Stance detection has been addressed as a task in its own right, where models have been developed based on data from
the Fake News Challenge \cite{riedel2017simple,thorne-EtAl:2017:NLPmJ,NAACL2018:stance,hanselowski-EtAl:2018:C18-1}, or from
SemEval-2017 Task~8 \cite{derczynski-EtAl:2017:SemEval,dungs-EtAl:2018:C18-1,ZubiagaKLPLBCA18}.
It has also been studied for other languages such as Arabic \cite{DarwishMZ17,baly-EtAl:2018:N18-2}.

\subsection{Source Reliability Estimation}

Unlike stance detection, the problem of source reliability remains largely under-explored.
In the case of social media, it concerns modeling the user\footnote{User modeling in social media and news community forums has focused on finding malicious users such as opinion manipulation \emph{trolls}, paid \cite{Mihaylov2015ExposingPO} or just perceived \cite{Mihaylov2015FindingOM,mihaylov-nakov:2016:P16-2,InternetResearchJournal:2018,AAAI2018:factchecking}, \emph{sockpuppets} \cite{Maity:2017:DSS:3022198.3026360}, \emph{Internet water army} \cite{Chen:2013:BIW:2492517.2492637}, and \emph{seminar users} \cite{SeminarUsers2017}.} who posted a particular message/tweet, while in the case of the Web, it is about the trustworthiness of the source (the URL domain, the medium).
The latter is our focus in this paper.

In previous work, the source reliability of news media has often been estimated automatically based on the general stance of the target medium with respect to known manually fact-checked claims, without access to gold labels about the overall medium-level factuality of reporting \cite{mukherjee2015leveraging,popat2016credibility,Popat:2017:TLE:3041021.3055133,Popat:2018:CCL:3184558.3186967}.
The assumption is that reliable media agree with true claims and disagree with false ones, while for unreliable media it is mostly the other way around.
The trustworthiness of Web sources has also been studied from a Data Analytics perspective.
For instance, \citet{Dong:2015:KTE:2777598.2777603} proposed that a trustworthy source is one that contains very few false facts.
In this paper, we follow a different approach by studying the source reliability as a task in its own right, using manual gold annotations specific for the task.

Note that estimating the reliability of a source is important not only when fact-checking a claim \cite{Popat:2017:TLE:3041021.3055133,DBLP:conf/aaai/NguyenKLW18}, but it also gives an important prior when solving article-level tasks such as ``fake news'' and click-bait detection \cite{brill2001online,finberg2002digital,Hardalov2016,RANLP2017:clickbait,desarkar-yang-mukherjee:2018:C18-1,Pan:KG:2018,prezrosas-EtAl:2018:C18-1}.

\subsection{``Fake News'' Detection}

Most work on ``fake news'' detection has relied on medium-level labels, which were then assumed to hold for all articles from that source.

\citet{DBLP:journals/corr/HorneA17} analyzed three small datasets ranging from a couple of hundred to a few thousand articles from a  couple of dozen sources, comparing (\emph{i})~real news vs. (\emph{ii})~``fake news'' vs. (\emph{iii})~satire, and found that the latter two have a lot in common across a number of dimensions. They designed a rich set of features that analyze the text of a news article, modeling its complexity, style, and psychological characteristics. They found that ``fake news'' pack a lot of information in the title (as the focus is on users who do not read beyond the title), and use shorter, simpler, and repetitive content in the body (as writing fake information takes a lot of effort). Thus, they argued that the title and the body should be analyzed separately.

In follow-up work, \citet{DBLP:journals/corr/abs-1803-10124} created a large-scale dataset covering 136K articles from 92 sources from \url{opensources.co}, which they characterize based on 130 features from seven categories: structural, sentiment, engagement, topic-dependent, complexity, bias, and morality. We use this set of features when analyzing news articles.

In yet another follow-up work, \citet{Horne:2018:ANL:3184558.3186987} trained a classifier to predict whether a given news article is coming from a reliable or from an unreliable (``\emph{fake news}'' or \emph{conspiracy})\footnote{We show in parentheses, the labels from \url{opensources.co} that are used to define a category.} source. Note that they assumed that all news from a given website would share the same reliability class. Such an assumption is fine for training (distant supervision), but we find it problematic for testing, where we believe manual documents-level labels are needed. 

\citet{DBLP:journals/corr/PotthastKRBS17} used 1,627 articles from nine sources, whose factuality has been manually verified by professional journalists from BuzzFeed. They applied stylometric analysis, which was originally designed for authorship verification, to predict factuality (fake vs. real).

\citet{rashkin-EtAl:2017:EMNLP2017} focused on the language used by ``fake news'' and compared the prevalence of several features in articles coming from trusted sources vs. hoaxes vs. satire vs. propaganda. However, their linguistic analysis and their automatic classification were at the article level and they only covered eight news media sources.

\noindent Unlike the above work, (\emph{i)}~we perform classification at the news medium level rather than focusing on an individual article. Thus, (\emph{ii})~we use reliable manually-annotated labels as opposed to noisy labels resulting from projecting the category of a news medium to all news articles published by this medium (as most of the work above did).\footnote{Two notable exceptions are \cite{DBLP:journals/corr/PotthastKRBS17} and \cite{prezrosas-EtAl:2018:C18-1}, who use news articles whose factuality has been manually checked and annotated.} Moreover, (\emph{iii})~we use a much larger set of news sources, namely 1,066, which is 1-2 orders of magnitude larger than what was used in previous work. Furthermore, (\emph{iv})~we use a larger number of features and a wider variety of feature types compared to the above work, including features extracted from knowledge sources that have been largely neglected in the literature so far such as information from Wikipedia and the structure of the medium's URL.

\subsection{Media Bias Detection}

As we mentioned above, bias was used as a feature for ``fake news'' detection \cite{DBLP:journals/corr/abs-1803-10124}.
It has also been the target of classification, e.g.,~\citet{Horne:2018:ANL:3184558.3186987} predicted whether an article is biased (\emph{political} or \emph{bias}) vs. unbiased. 
Similarly, \citet{DBLP:journals/corr/PotthastKRBS17} classified the bias in a target article as
(\emph{i})~left vs. right vs. mainstream, or as
(\emph{ii})~hyper-partisan vs. mainstream. 
Finally, \citet{rashkin-EtAl:2017:EMNLP2017} studied propaganda, which can be seen as extreme bias.
See also a recent position paper \cite{Pitoura:2018:MBO:3186549.3186553} and an overview on bias the Web \cite{Baeza-Yates:2018:BW:3229066.3209581}.

Unlike the above work, we focus on bias at the medium level rather than at the article level. Moreover, we work with fine-grained labels on an ordinal scale rather then having a binary setup (some work above had three degrees of bias, while we have seven).

\section{Method}
\label{sec:method}

In order to predict the factuality of reporting and the bias for a given news medium, we collect information from multiple relevant sources, which we use to train a classifier.
In particular, we collect a rich set of features derived from
(\emph{i})~a sample of articles from the target news medium,
(\emph{ii})~its Wikipedia page if it exists,
(\emph{iii})~its Twitter account if it exists,
(\emph{iv})~the structure of its URL, and
(\emph{v})~information about the Web traffic it has attracted.
We describe each of these sources below.

\paragraph{Articles}
We argue that analysis (textual, syntactic and semantic) of the content of the news articles published by a given target medium should be critical for assessing the factuality of its reporting, as well as of its potential bias.
Towards this goal, we borrow a set of 141 features that were previously proposed for detecting ``fake news'' articles \cite{DBLP:journals/corr/abs-1803-10124}, as we have described above.
These features are used to analyze the following article characteristics:
\begin{itemize}
\item \textbf{Structure}: POS tags, linguistic features based on the use of specific words (function words, pronouns, etc.), and features for clickbait title classification from \cite{clickbait:2016};
\item \textbf{Sentiment}: sentiment scores using lexicons \citep{recasens2013linguistic,mitchell2013geography} and full systems \cite{gilbert2014vader};
\item \textbf{Engagement}: number of shares, reactions, and comments on Facebook;
\item \textbf{Topic}: lexicon features to differentiate between science topics and personal concerns;
\item \textbf{Complexity}: type-token ratio, readability, number of cognitive process words (identifying discrepancy, insight, certainty, etc.);
\item \textbf{Bias}: features modeling bias using lexicons~\citep{recasens2013linguistic,mukherjee2015leveraging} and subjectivity as calculated using pre-trained classifiers~\cite{horne2017identifying};
\item \textbf{Morality}: features based on the Moral Foundation Theory \cite{graham2009liberals} and lexicons \cite{lin2017acquiring}
\end{itemize}

Further details are available in~\cite{DBLP:journals/corr/abs-1803-10124}.
For each target medium, we retrieve some articles, then we calculate these features separately for the title and for the body of each article, and finally we average the values of the 141 features over the set of retrieved articles.

\paragraph{Wikipedia}
We further leverage Wikipedia as an additional source of information that can help predict the factuality of reporting and the bias of a target medium.
For example, the absence of a Wikipedia page may indicate that a website is not credible.
Also, the content of the page might explicitly mention that a certain website is satirical, left-wing, or has some property related to our task.

\noindent Accordingly, we extract the following features:
\begin{itemize}
\item {\it Has Page:} indicates whether the target medium has a Wikipedia page;
\item Vector representation for each of the following segments of the Wikipedia page, whenever applicable:
{\it Content},
{\it Infobox},
{\it Summary},
{\it Categories},
and
{\it Table of Contents}.
We generate these representations by averaging the word embeddings (pretrained word2vec embeddings) of the corresponding words.
\end{itemize}

\paragraph{Twitter}
Given the proliferation of social media, most news media have Twitter accounts, which they use to reach out to more users online.
The information that can be extracted from a news medium's Twitter profile can be valuable for our tasks.
In particular, we use the following features:
\begin{itemize}
\item {\it Has Account:} Whether the medium has a Twitter account. We check this based on the top results for a search against Google, restricting the domain to \url{twitter.com}. The idea is that media that publish unreliable information might have no Twitter accounts.
\item {\it Verified:} Whether the account is verified by Twitter. The assumption is that ``fake news'' media would be less likely to have their Twitter account verified. They might be interested in pushing their content to users via Twitter, but they would also be cautious about revealing who they are (which is required by Twitter to get them verified).
\item {\it Created:} The year the account was created. The idea is that accounts that have been active over a longer period of time are more likely to belong to established media. 
\item {\it Has Location:} Whether the account provides information about its location. The idea is that established media are likely to have this public, while ``fake news'' media may want to hide it.
\item {\it URL Match:} Whether the account includes a URL to the medium, and whether it matches the URL we started the search with. Established media are interested in attracting traffic to their website, while fake media might not. Moreover, some fake accounts mimic genuine media, but have a slightly different domain, e.g.,~\texttt{.com.co} instead of \texttt{.com}.
\item {\it Counts}: Statistics about the number of friends, statuses, and favorites. Established media might have higher values for these.
\item {\it Description:} A vector representation generated by averaging the \emph{Google News} embeddings \cite{mikolov-yih-zweig:2013:NAACL-HLT} of all words of the profile description paragraph. These short descriptions might contain an open declaration of partisanship, i.e., left or right political ideology (bias). This could also help predict factuality as extreme partisanship often implies low factuality. 
In contrast, ``fake news'' media might just leave this description empty, while high-quality media would want to give some information about who they are.
\end{itemize}

\begin{table*}[tbh]
\centering
\scalebox{0.775}{
\begin{tabular}{l|l|l|l|l}
\toprule
\bf Name & \bf URL & \bf Factuality & \bf Twitter Handle & \bf Wikipedia page \\ \midrule
Associated Press & \url{http://apnews.com} & $^{\star}$Very High & @apnews & \url{~/wiki/Associated_Press} \\
NBC News & \url{http://www.nbcnews.com/} & High & @nbcnews & \url{~/wiki/NBC_News} \\
Russia Insider & \url{http://russia-insider.com} & Mixed & @russiainsider & \url{~/wiki/Russia_Insider} \\
Patriots Voice & \url{http://patriotsvoice.info/} & Low & @pegidaukgroup & N/A \\
\bottomrule
\end{tabular}}
\caption{Examples of media with various factuality scores. \small{($^{\star}$In our experiments, we treat \emph{Very High} as \emph{High}.)}\label{tab:example:factuality}}
\end{table*}

\begin{table*}[tbh]
\centering
\scalebox{0.78}{
\begin{tabular}{ l|l|l|l|l}
\toprule
\bf Name & \bf URL & \bf Bias & \bf Twitter Handle & \bf Wikipedia page \\ \midrule
Loser.com & \url{http://loser.com} & Extreme Left & @Loser\_dot\_com & \url{~/Loser.com} \\
Die Hard Democrat & \url{http://dieharddemocrat.com/} & Left & @democratdiehard & N/A \\
Democracy 21 & \url{http://www.democracy21.org/} & Center-Left & @fredwertheimer & \url{~/Democracy_21} \\
Federal Times & \url{http://www.federaltimes.com/} & Center & @federaltimes & \url{~/Federal_Times} \\
Gulf News & \url{http://gulfnews.com/} & Center-Right & @gulf\_news & \url{~/Gulf\_News} \\
Fox News & \url{http://www.foxnews.com/} & Right & @foxnews & \url{~/Fox\_News} \\
Freedom Outpost & \url{http://freedomoutpost.com/} & Extreme Right & @FreedomOutpost & N/A \\
\bottomrule
\end{tabular}}
\caption{Examples of media with various bias scores.\label{tab:example:bias}}
\end{table*}

\paragraph{URL}
We also collect additional information from the website's URL using character-based modeling and hand-crafted features.
URL features are commonly used in phishing website detection systems to identify malicious URLs that aim to mislead users~\cite{ma2009identifying}.
As we want to predict a website's factuality, using URL features is justified by the fact that low-quality websites sometimes try to mimic popular news media by using a URL that looks similar to the credible source. We use the following URL-related features: 
\begin{itemize}
\item {\it Character-based:} Used to model the URL by representing it in the form of a one-hot vector of character {\it n}-grams, where $n\in[2,5]$. Note that these features are not used in the final system as they could not outperform the baseline (when used in isolation). 
\item {\it Orthographic:} These features are very effective for detecting phishing websites, as malicious URLs tend to make excessive use of special characters and sections, and ultimately end up being longer.
For this work, we use the length of the URL, the number of sections and the excessive use of special characters such as digits, hyphens and dashes.
In particular, we identify whether the URL contains digits, dashes or underscores as individual symbols, which were found to be useful as features for detecting phishing URLs~\cite{basnet2014learning}.
We also check whether the URL contains short (less than three symbols) or long sections (more than ten symbols), as a high number of such sections could indicate an irregular URL.
\item {\it Credibility:} Model the website's URL credibility by analyzing whether it
({\it i})~uses \texttt{https://},
({\it ii})~resides on a blog-hosting platform such as \url{blogger.com}, and
({\it iii})~uses a special top-level domain, e.g.,~\texttt{.gov} is for governmental websites, which are generally credible and unbiased, whereas \texttt{.co} is often used to mimic \texttt{.com}. 
\end{itemize}

\paragraph{Web Traffic}
Analyzing the web traffic to the website of the medium might be useful for detecting phishy websites that come and disappear in certain patterns.
Here, we only use the reciprocal value of the website's {\it Alexa Rank},\footnote{\url{http://www.alexa.com/}} which is a global ranking for over 30 million websites in terms of the traffic they receive.

We evaluate the above features in Section~\ref{sec:experiments}, both individually and as groups, in order to determine which ones are important to predict factuality and bias, and also to identify the ones that are worth further investigation in future work.

\section{Experiments and Evaluation}\label{sec:experiments}

\subsection{Data}\label{subsec:data}
We use information about news media listed on the Media Bias/Fact Check (MBFC) website,\footnote{\url{https://mediabiasfactcheck.com}} which contains manual annotations and analysis of the factuality of reporting and/or bias for over 2,000 news websites.
Our dataset includes 1,066 websites for which \emph{both} bias and factuality labels were explicitly provided, or could be easily inferred (e.g.,~\emph{satire} is of low factuality).

\noindent We model factuality on a 3-point scale (\emph{Low}, \emph{Mixed}, and \emph{High}),\footnote{MBFC also uses \emph{Very High} as a label, but due to its very small size, we merged it with \emph{High}.} and bias on a 7-point scale ({\emph{Extreme-Left}, \emph{Left}, \emph{Center-Left}, \emph{Center}, \emph{Center-Right}, \emph{Right}, and {\it Extreme-Right}).

Some examples from our dataset are presented in Table~\ref{tab:example:factuality} for factuality  of reporting, and in Table~\ref{tab:example:bias} for bias.
In both tables, we show the names of the media, as well as their corresponding Twitter handles and Wikipedia pages, which we found automatically.
Overall, 64\% of the websites in our dataset have Wikipedia pages, and 94\% have Twitter accounts.
In cases of ``fake news'' sites that try to mimic real ones, e.g.,~\url{ABCnews.com.co} is a fake version of \url{ABCnews.com}, it is possible that our Twitter extractor returns the handle for the real medium. This is where the \emph{URL Match} feature comes handy (see above).

Table~\ref{tab:stats} provides detailed statistics about the dataset.
Note that we have 1-2 orders of magnitude more media sources than what has been used in previous studies, as we already mentioned in Section~\ref{sec:related} above.

\begin{table}[tbh]
\centering
\small
\begin{tabular}{ll|lr}
\toprule
\multicolumn{2}{c|}{\bf Factuality} & \multicolumn{2}{c}{\bf Bias} \\ \midrule
Low & 256 & Extreme-Left & 21 \\
Mixed & 268 & Left & 168 \\
High & 542 & Center-Left & 209 \\
 & & Center & 263 \\
 & & Center-Right & 92 \\
 & & Right & 157 \\
 & & Extreme-Right & 156 \\ \bottomrule
\end{tabular}
\caption{Label distribution (counts) in our dataset.\label{tab:stats}}
\end{table}

\begin{table*}[tbh]
\centering
\scalebox{0.85}{
\begin{tabular}{l|l|c||cccc||cccc}
\toprule
\bf Source & \bf Feature & \bf Dim. & \multicolumn{4}{c||}{\bf Factuality} & \multicolumn{4}{c}{\bf Bias} \\
 & & & \bf Macro-F$_1$ & \bf Acc. & \bf MAE & \bf MAE$^M$ & \bf Macro-F$_1$ & \bf Acc. & MAE & MAE$^M$ \\ \midrule
\multicolumn{3}{l||}{Majority Baseline}                 & 22.47	& 50.84 & 0.73 & 1.00 &    5.65 & 24.67 & 1.39 & 1.71 \\ \midrule
Traffic & \it Alexa rank                        & 1     & 22.46 & 50.75 & 0.73 & 1.00 &    7.76 & 25.70 & 1.38 & 1.71 \\ \midrule
URL & \it URL structure                         & 12    & 39.30 & 53.28 & 0.68 & 0.81 &   13.50 & 23.64 & 1.65 & 2.06 \\ \midrule
\multirow{8}{*}{Twitter} & 
   \it created at.                              & 1     & 30.72 & 52.91 & 0.69 & 0.92 &    5.65 & 24.67 & 1.39 & 1.71 \\
 & \it has account                              & 1     & 30.72 & 52.91 & 0.69 & 0.92 &    5.65 & 24.67 & 1.39 & 1.71 \\
 & \it verified                                 & 1     & 30.72 & 52.91 & 0.69 & 0.92 &    5.65	& 24.67 & 1.39 & 1.71 \\
 & \it has location                             & 1     & 36.73 & 52.72 & 0.69 & 0.82 &    9.44 & 24.86 & 1.54 & 1.85 \\
 & \it URL match                                & 2     & 39.98 & 54.60 & 0.66 & 0.72 &   10.16 & 25.61 & \bf 1.51 & 1.97 \\
 & \it description                              & 300   & 44.79 & 51.41 & 0.65 & 0.70 &   19.08 & 25.33 & 1.73 & 2.04 \\
 & \it counts                                   & 5     & 46.88 & \bf57.22 & \bf0.57 & 0.66 &   18.34 & 24.86 & 1.62 & 2.01 \\
 & \it Twitter -- All                           & 308   & \bf48.23 & 54.78 & 0.59 & \bf0.64 &   \bf 21.38 & \bf 27.77 & 1.58 & \bf 1.83 \\ \midrule
\multirow{7}{*}{Wikipedia} & 
   \it has page                                 & 1     & 43.53 & 59.10 & 0.57 & 0.63 &   14.33 & 26.83 & 1.63 & 2.14 \\
 & \it table of content                         & 300   & 43.95 & 51.04 & 0.60 & 0.65 &   15.10 & 22.96 & 1.86 & 2.25 \\
 & \it categories                               & 300   & 46.36 & 53.70 & 0.65 & 0.61 &   25.64 & 32.16 & 1.70 & 2.10 \\
 & \it information box                          & 300   & 46.39 & 51.14 & 0.71 & 0.65 &   19.79 & 26.85 & 1.68 & 1.99 \\
 & \it summary                                  & 300   & 51.88 & 58.91 & 0.54 & 0.52 &   30.02	& 37.43 & 1.47 & \bf 1.98 \\
 & \it content                                  & 300   & 55.29 & 62.10 & 0.51 & 0.50 &   \bf 30.92 & \bf 38.61 & \bf 1.51 & 2.01 \\
 & \it Wikipedia -- All                         & 301   & \bf55.52 & \bf62.29 & \bf0.50 & \underline{\bf0.49} &   28.66 & 35.93  & 1.51 & 2.00\\ \midrule
\multirow{3}{*}{Articles} & 
   \it title                                    & 141  & 53.20 & 59.57 & 0.51 & 0.58 &   30.91 & 37.52 & 1.29 & 1.53 \\
 & \it body                                     & 141  & \underline{\bf 58.02} & \underline{\bf 64.35} & \underline{\bf 0.43} & \bf 0.51 &   \underline{\bf 36.63} & \underline{\bf 41.74} & \underline{\bf 1.15} & \underline{\bf 1.43} \\ \bottomrule
\end{tabular}}
\caption{Results for factuality and bias prediction.
{\bf Bold} values indicate the best-performing feature type in its family of features, while \underline{\bf underlined} values indicate the best-performing feature type overall.
\label{tab:results}}
\end{table*}

\noindent In order to compute the article-related features, we did the following:
(\emph{i})~we crawled 10--100 articles per website (a total of 94,814),
(\emph{ii})~we computed a feature vector for each article, and
(\emph{iii})~we averaged the feature vectors for the articles from the same website to obtain the final vector of article-related features.

\subsection{Experimental Setup}
\label{subsec:setup}

We used the above features in a Support Vector Machine (SVM) classifier, training a separate model for factuality and for bias.
We report results for 5-fold cross-validation.
We tuned the SVM hyper-parameters, i.e., the cost $C$, the kernel type, and the kernel width $\gamma$, using an internal cross-validation on the training set and optimizing macro-averaged $F_1$. Generally, the RBF kernel performed better than the linear kernel.

We report accuracy and macro-averaged $F_1$ score. 
We also report Mean Average Error (MAE), which is relevant given the ordinal nature of both the factuality and the bias classes, and also MAE$^M$, which is a variant of MAE that is more robust to class imbalance.
See \cite{Baccianella:2009qd,rosenthal-farra-nakov:2017:SemEval} for more details about MAE$^{M}$ vs. MAE.

\subsection{Results and Discussion}
\label{subsec:results}
We present in Table~\ref{tab:results} the results of using features from the different sources proposed in Section~\ref{sec:method}.
We start by describing the contribution of each feature type towards factuality and bias.

We can see that the textual features extracted from the {\sc Articles} yielded the best performance on factuality.
They also perform well on bias, being the only type that beats the baseline on MAE.
These results indicate the importance of analyzing the contents of the target website.
They also show that using the \emph{titles} only is not enough, and that the article \emph{bodies} contain important information that should not be ignored.

\begin{table*}[tbh]
\centering
\scalebox{0.95}{
\begin{tabular}{l|cccc}
\toprule
\bf Features & \bf Macro-F$_1$ & \bf Acc. & \bf MAE & \bf MAE$^M$ \\ \midrule
\sc Majority Baseline   & 22.47	& 50.84 & 0.73 & 1.00 \\ 
\sc Full                & 59.91 & 65.48 & 0.41 & 0.44 \\ \midrule
\sc Full w/o Traffic    & 59.90 & 65.39 & 0.41 & 0.43 \\
\sc Full w/o Twitter    & 59.52 & 65.10 & 0.41 & 0.47 \\
\sc Full w/o URL        & 57.23 & 63.32 & 0.44 & 0.49 \\
\sc Full w/o Articles	& 56.15 & 63.13 & 0.46 & 0.51 \\
\sc Full w/o Wikipedia  & 55.93 & 63.23 & 0.44 & 0.52 \\
\bottomrule
\end{tabular}}
\caption{\label{tab:ablation_factuality}Ablation study for the contribution of each feature type for predicting the factuality of reporting.}

\bigskip

\scalebox{0.94}{
\begin{tabular}{l||cccc||cccc}
\toprule
\bf Features & \multicolumn{4}{c||}{\bf 7-Way Bias} & \multicolumn{4}{c}{\bf 3-Way Bias} \\
 & \bf Macro-F$_1$ & \bf Acc. & \bf MAE & \bf MAE$^M$ & \bf Macro-F$_1$ & \bf Acc. & \bf MAE & \bf MAE$^M$ \\ \midrule
\sc Majority Baseline   &  5.65 & 24.67 & 1.39 & 1.71 &   22.61 & 51.33 & 0.49 & 0.67 \\
\sc Full                & 37.50 & 39.87 & 1.25 & 1.55 &   61.31 & 68.86 & 0.39 & 0.53 \\ \midrule
\sc Full w/o Traffic    & 37.49 & 39.84 & 1.25 & 1.55 &   61.30 & 68.86 & 0.38 & 0.53 \\
\sc Full w/o Twitter    & 36.88 & 39.49 & 1.20 & 1.38 &   63.27 & 69.89 & 0.38 & 0.50 \\
\sc Full w/o URL        & 36.60 & 39.68 & 1.24 & 1.48 &   60.93 & 68.11 & 0.40 & 0.53 \\
\sc Full w/o Wikipedia  & 34.75 & 37.62 & 1.33 & 1.58 &   59.92 & 66.89 & 0.41 & 0.54 \\
\sc Full w/o Articles	& 29.95 & 36.96 & 1.40 & 1.85 &   53.67 & 62.48 & 0.47 & 0.62 \\ \bottomrule
\end{tabular}}
\caption{Ablation study for the contribution of each feature type for predicting media bias.\label{tab:ablation_bias}}
\end{table*}

Overall, the {\sc Wikipedia} features are less useful for factuality, and perform reasonably well for bias.
The best features from this family are those about the page \emph{content}, which includes a general description of the medium, its history, ideology and other information that can be potentially helpful.
Interestingly, the \emph{has page} feature alone yields sizable improvement over the baseline, especially for factuality.
This makes sense given that trustworthy websites are more likely to have Wikipedia pages; yet, this feature does not help much for predicting political bias.

\noindent The {\sc Twitter} features perform moderately for factuality and poorly for bias.
This is not surprising, as we normally may not be able to tell much about the political ideology of a website just by looking at its Twitter profile (not its tweets!) unless something is mentioned in its \emph{description}, which turns out to perform better than the rest of the features from this family.
We can see that the {\it has twitter} feature is less effective than {\it has wiki} for factuality, which makes sense given that Twitter is less regulated than Wikipedia.
Note that the {\it counts} features yield reasonable performance, indicating that information about activity (e.g.,~number of statuses) and social connectivity (e.g.,~number of followers) is useful.
Overall, the {\sc Twitter} features seem to complement each other, as their union yields the best performance on factuality.

The {\sc URL} features are better used for factuality rather than bias prediction.
This is mainly due to the nature of these features, which are aimed at detecting phishing websites, as we mentioned in Section~\ref{sec:method}.
Overall, this feature family yields slight improvements, suggesting that it can be useful when used together with other features.

Finally, the {\it Alexa rank} does not improve over the baseline, which suggests that more sophisticated {\sc Traffic}-related features might be needed.

\subsection{Ablation Study}

Finally, we performed an ablation study in order to evaluate the impact of removing one family of features at a time, as compared to the {\sc Full} system, which uses all the features.
We can see in Tables~\ref{tab:ablation_factuality} and~\ref{tab:ablation_bias} that the {\sc Full} system achieved the best results for factuality, and the best macro-F$_1$ for bias, suggesting that the different types of features are largely complementary and capture different aspects that are all important for making a good classification decision.

For factuality, excluding the {\sc Wikipedia} features yielded the biggest drop in performance.
This suggests that they provide information that may not be available in other sources, including the {\sc Articles}, which achieved better results alone.
On the other hand, excluding the {\sc Traffic} feature had no effect on the model's performance.

For bias, we experimented with classification on both a 7-point and a 3-point scale.\footnote{We performed the following mapping:\\ \{{\it Extreme-Right, Right}\}$\to$Right, \{{\it Extreme-Left, Left}\}$\to$Left, and \{{\it Center, Right-Center, Left-Center}\}$\to$Center}
Similarly to factuality, the results in Table~\ref{tab:ablation_bias} indicate that {\sc Wikipedia} offers complementary information that is critical for bias prediction, while {\sc Traffic} makes virtually no difference.

\section{Conclusion and Future Work}
\label{sec:conclusion}

We have presented a study on predicting factuality of reporting and bias of news media, focusing on characterizing them as a whole. 
These are under-studied, but arguably important research problems, both in their own right and as a prior for fact-checking systems.

We have created a new dataset of news media sources that has annotations for both tasks and is 1-2 orders of magnitude larger than what was used in previous work. We are releasing the dataset and our code, which should facilitate future research.

We have experimented with a rich set of features derived from the contents of (\emph{i})~a sample of articles from the target news medium, (\emph{ii})~its Wikipedia page, (\emph{iii})~its Twitter account, (\emph{iv})~the structure of its URL, and (\emph{v})~information about the Web traffic it has attracted.
This combination, as well as some of the types of features, are novel for this problem.

Our evaluation results have shown that most of these features have a notable impact on performance, with the articles from the target website, its Wikipedia page, and its Twitter account being the most important (in this order).
We further performed an ablation study of the impact of the individual types of features for both tasks, which could give general directions for future research.

In future work, we plan to address the task as ordinal regression, and further to model the inter-dependencies between factuality and bias in a joint model. 
We are also interested in characterizing the factuality of reporting for media in other languages. Finally, we want to go beyond \emph{left vs. right} bias that is typical of the Western world and to model other kinds of biases that are more relevant for other regions, e.g.,~\emph{islamist vs. secular} is one such example for the Muslim World.

\section*{Acknowledgments}

This research was carried out in collaboration between the MIT Computer Science and Artificial Intelligence Laboratory (CSAIL) and the Qatar Computing Research Institute (QCRI), HBKU.

We would like to thank Israa Jaradat, Kritika Mishra, Ishita Chopra, Laila El-Beheiry, Tanya Shastri, and Hamdy Mubarak for helping us with the data extraction, cleansing, and preparation.

Finally, we thank the anonymous reviewers for their constructive comments, which have helped us improve this paper.

\bibliography{bibliography}
\bibliographystyle{acl_natbib_nourl}

\end{document}